\documentclass[12pt,preprint]{aastex}

\newcommand{\bb}{\begin{equation}}
\newcommand{\ee}{\end{equation}}

\shorttitle{Stellar magnetic flux tubes}
\shortauthors{Rajaguru et al.}

\begin{document}

\title{Convective intensification of magnetic flux tubes in stellar photospheres}
\author{S. P. Rajaguru\altaffilmark{1}, R. L. Kurucz\altaffilmark{2} and S. S. Hasan\altaffilmark{1}}
\altaffiltext{1}{Indian Institute of Astrophysics, Bangalore, India.}
\altaffiltext{2}{Center for Astrophysics,
    60 Garden Street, Cambridge, MA 02138, USA}

\begin{abstract} The convective collapse of thin magnetic flux tubes in the photospheres of
sun-like stars is investigated using realistic models of the superadiabatic upper convection
zone layers of these stars.  The strengths of convectively stable flux tubes are computed as a
function of surface gravity and effective temperature. We find that while stars with
T$_{eff}\ge$ 5500 K and log $g$$\ge$ 4.0 show flux tubes highly evacuated of gas, and hence
strong field strengths, due to convective collapse, cooler stars exhibit flux tubes with lower
field strengths.  Observations reveal the existence of field strengths close to thermal
equipartition limits even in cooler stars, implying highly evacuated tubes, for which we
suggest possible reasons.  \end{abstract}


\section{Introduction}
\label{sec:intro}

An important physical quantity basic to the understanding of stellar magnetism is the field
strength in small-scale flux tubes and spots on the photospheres. Measurements and
interpretation of magnetic field strengths \citep{saar88,saar96,guenther97} in cool
main-sequence stars other than the Sun have been a subject of much debate recently
\citep{basrietal90, safier99}, because of the difficulties associated with the modeling of the
atmospheric structural changes that the inhomogeneous fibril state of the magnetic field
introduces \citep{basrietal90}. However, recent improvements in observing techniques
\citep{saar96,donatietal97,j.krulletal99} have increased the reliability of observational
determination of stellar magnetic field strengths. The observed Zeeman broadening on cool stars
is believed to be produced by small flux tubes that appear bright, similar to the solar
magnetic network and facular bright points, rather than spots, which are much fainter than
quiet photospheres and hence contribute little to the stellar profiles
\citep{basrietal90,safier99}.

Based on measurements of Zeeman effect in cool main sequence stars \citet{saar94,saar96} has
inferred that the magnetic field strengths are close to the thermal equipartition field
strengths B$_{eq}$=$\sqrt{ 8\pi p_{e}}$ at the observed levels in the atmosphere, where $p_{e}$
is the gas pressure in the unmagnetised atmosphere, with a conclusion that the surface
distribution is in the form of highly evacuated small flux tubes in pressure balance with the
ambient atmosphere. Various other measurements \citep{saarlinsky85,basri-marcy94,j.krulletal99}
also establish such field strengths leading to the general acceptance \citep{linsky99} that
stellar surface field strengths scale as the square root of the surface gas pressure. We note
that B$_{eq}$, at any geometrical level in a stellar atmosphere, is the maximum possible field
strength for flux tubes confined by gas pressure and can be easily determined
\citep{bunte-saar93,safier99} as a function of log $g$ and T$_{eff}$ using atmospheric models
(e.g., \citet{kurucz93}), where $g$ and T$_{eff}$ denote the surface gravity and effective
temperature.

The widely accepted mechanism to account for the kG range field strengths in solar magnetic
flux tubes is the superadiabatic effect or convective collapse (CC, hereafter) 
\citep{parkr78,sptz79}.  In this Letter, we report results from a study of this mechanism
in the stellar context and test its efficiency as a function of log $g$ and T$_{eff}$.

Concentration of magnetic fields into discrete flux tubes or sheaths has long been recognised
as a general consequence of the interaction between magnetic fields and cellular convection
that expels the flux into downdrafts
\citep{parkr63,galloway-weiss81,proctor-weiss82}.  High-resolution observations of solar
surface magnetic fields confirm the operation of such flux expulsion process. The CC,
which is a consequence of
thermal insulation of the expelled magnetic flux against convective motions and the
superadiabatic thermal stratification of the ambient atmosphere, further intensifies the field
to observed values \citep{parkr78}. It is a convective instability, modified by the magnetic
field, and drives a down flow along the field lines of a flux concentration leading to the
evacuation of gas inside; lateral pressure balance ensures that a highly compressed intense
field flux tube or sheath is formed. This instability develops very rapidly, typically on a
free-fall time scale, and is faster than other MHD instabilities present.  Hence, the criterion
for convective stability can be regarded as a test to check for the occurrence of flux tubes in
stars.  On the Sun, this instability has a typical growth time of about 2 - 5 minutes
\citep{hasan84,raj00}. Other important instabilities are the interchange instability
\citep{parkr75,piddn75} connected with the field topology, which has been studied in the
stellar context by \citet{bunte-saar93}, and the Kelvin- Helmholtz instability
\citep{schuss79,tsinga80} connected with the relative motion between the field confined and the
surrounding gas.

We use model atmospheres constructed using the ATLAS9 model atmosphere code \citep{kurucz93},
which can extend the \citet{kurucz93} atmospheres to cover deeper regions of the convection
zone, to study the CC of thin flux tubes embedded in them. The critical field strengths for the
tubes to be stable, which indicate the amount of evacuation that stellar flux tubes undergo as
a result of the collapse process, are found.

\section{Convective Stability of Stellar Thin Flux Tubes}
\label{sec:main}

A simple and very effective modeling of the dynamics of thin flux tubes, including the
important effects of stratification and compressibility, is provided by the thin flux tube
approximation \citep{robweb78} to the MHD equations. The thin flux tube equations have proved
very useful in the study of small-scale solar magnetic elements and have been successful in
explaining a variety of observed phenomena (Solanki~1993 and references therein). We solve the
same second order differential equation derived by \citet{sptz79} based on linear stability
analysis of thin tube equations in the adiabatic limit, which is valid for arbitrary background
stratifications, employing a grid of stellar model atmospheres. We do not reproduce the
equation here, and we refer the reader to \citet{sptz79} for details.  The stability of a flux
tube is controlled by $\beta$, which is the ratio of gas to magnetic pressure inside the tube.
Weak field (i.e, high $\beta$) flux concentrations that are formed at the downdrafts of
convective cells are convectively unstable and collapse until their field strengths reach the
critical value at which the instability is quenched. Thus the instability is self-limiting and
leads to a lower energy stable equilibrium state for flux tubes \citep{spruit79}.

\subsection{Stellar Model Atmospheres}
\label{sec:models}

Models of stellar convective envelopes that extend downward to depths required to study the CC
have been specially constructed, extending the Kurucz (1993) models to cover deeper layers
using the ATLAS9 model atmosphere code. The convective flux is calculated using mixing length
theory. We computed models, with solar metallicity, by extrapolating the shallower models
downward one or two points, reconverging, and then repeating the process until a sufficient
depth in the atmosphere was obtained. Though the superadiabatic regions below the stellar
surfaces are usually thin, a study of the development of convectively driven motions in flux
tubes requires a substantial depth extent downward into the convection zone. 
For the sun, a depth extent of at least 5000 km is needed for the stability limit to
be independent of mechanical conditions at the boundary locations \citep{raj00}. For stars with
log($g$) $\ge$ 4.5, the ATLAS9 program fails to construct reliable models of sub-surface
regions upto depths that are required in our study. In those situations we attach adiabatic
polytropic models that smoothly match with the models from ATLAS9; we note however that, in
these cases, the realistic models from ATLAS9 cover the superadiabatic thin regions that drive
the convective instability in the flux tubes. The further extension with a polytropic
stratification is needed only to avoid breaking artificially the inertia of the convectively
driven down flow from the superadiabatic region above at the location where the shallower
ATLAS9 model ends. The depth extents, and two important quantities characterizing
the surface layers, viz., superadiabaticity $\delta = \nabla - \nabla_{ad}$, where $\nabla$ and
$\nabla_{ad}$ are the actual and adiabatic temperature gradients respectively, and adiabatic
index $\Gamma_{1}$, are tabulated in Table 1. The grid of models constructed have T$_{eff}$ in
the range 4000 -- 7000 K (in steps of 500 K) and log $g$ in the range 2.5 -- 5.0 (in steps of
0.5).  

\subsection{Convective Stability Limits} 
\label{sec:results}

We have numerically solved the linear eigenvalue problem (equations [6] and [8] of
\citet{sptz79}), which determines the growth rates 
and frequencies of the unstable modes and their eigenfunctions, for
stellar flux tubes using atmospheric models described in Section~\ref{sec:models}.  The
convective mode is the fastest growing and is identified from its eigenfunction which has no
node between the boundaries, corresponding to a monotonic downflow or upflow throughout the
extent of the tube. The critical values $\beta_{c}$ that determine stability against these
convective modes are found. Assumption of temperature equality between the flux tube and the
ambient medium makes $\beta$ depth independent \citep{sptz79}, and we evaluate the critical
field strengths from $\beta_{c}$ using values of $p_{e}$ at the height where the continuum
Rosseland optical depth $\tau_{c}$=1. The above is a good approximation close to the
photospheres, because the temperature differences between the flux tube and ambient medium are
minimal at these heights, as known from solar observations, and also the superadiabatic layers
that drive the convective instability are very close to the photosphere (Table 1).  The
critical field strengths are then given by B$_{c}$($g$,T$_{eff}$)=$\sqrt{8\pi
p_{e}/(1+\beta_{c})}$= f$_{cc}$($g$,T$_{eff}$)B$_{eq}$($g$,T$_{eff}$). The factor
f$_{cc}$=1/$\sqrt{(1+\beta_{c})}$ represents the efficiency of CC;  a value of f$_{cc}$=1,
i.e., $\beta_{c}$=0, corresponds to zero gas pressure, representing a fully evacuated flux
tube. The results are summarised in Figure 1. Shown are the critical field strengths
B$_{c}(\tau_{c}=1)$ (solid curve) required for convective stability and the maximum possible
field strengths B$_{eq}(\tau_{c}=1)$ (dashed curve), as a function of log $g$ for various
values of T$_{eff}$. The variation of B$_{eq}$ with log $g$ reflects the fact that the
photospheric gas pressure, for hydrostatic equilibrium, varies as $g$.  The dependence of
B$_{c}$ on log $g$ is modulated by the efficiency of CC, i.e., by f$_{cc}$($g$,T$_{eff}$).
Changing f$_{cc}$, i.e., varying efficiency of CC across the log $g$ -- T$_{eff}$ plane, would
determine the trends of B$_{c}$ that differ from B$_{eq}$ in Fig.~1.  To see this behaviour
clearly, we have also shown the variation of f$_{cc}$ (dotted curves) with log $g$ in Fig.~1.
Several trends are seen in the figure. The most striking feature is the increasing efficiency
of CC for log $g$$>$3.5 as T$_{eff}$ increases; the maximum efficiency occurrs at log $g$=4 and
T$_{eff}$=6500 K.  For log $g$ $<$ 3.5 all stars show inefficient collapse, although $\delta$
increases as log $g$ decreases for a fixed T$_{eff}$. Since gravity provides the main 
accelerating force on a displaced fluid element in a convectively unstable stratification, a
large reduction in gravity is expected to weaken the convective motions irrespective of
variations in $\delta$. This explains the inefficient CC at low values of log $g$.  For log
$g$$\ge$ 3.5, CC shows varying amount of dependencies on T$_{eff}$; the general feature is
decreasing efficiency of CC with T$_{eff}$. The largest differences between B$_{eq}$ and
B$_{c}$ are found for the coolest star considered (T$_{eff}$=4000 K). For example, for
T$_{eff}$=4000 K and log $g$=4.5, B$_{eq}$ and B$_{c}$ differ by about 1000 G. This inefficient
behaviour of CC finds an explanation in Figure 2, where we plot the maximum values of the
superadiabaticity $\delta$ against T$_{eff}$ for different values of log $g$; $\delta$
decreases monotonically with T$_{eff}$, for fixed log $g$, thus explaining the low efficiency
of CC in cooler stars.  As a general result, we find that CC is not efficient enough in stars
cooler than T$_{eff}$=5500 K, i.e., on the cooler side of the Sun in the main-sequence, owing
to the absence of significantly superadiabatic sub-surface thermal structure, to produce field
strengths close to B$_{eq}$.

We emphasise that the stability limits found in this study are from a global analsyis and with
effects of vertical gradients in ionization and adiabatic index taken into account.  There are
no known analytic criteria that capture the above effects. An analytic criterion derived by
\citet{robweb78} for finite extent vertical thin flux tubes is both necessary and sufficient
but for linear (polytropic) temperature profiles and do not include effects of gradients in
ionization and adiabatic index.  Despite its limitations we use it to compare and understand
qualititatively our results. This criterion yields critical values of $\beta$ given by, 
\bb
\beta_{c,RW}=\frac{\nabla^{2}}{2\delta}\left [\left (\frac{2\pi}{ln(H_{b}/H_{0})}\right )^{2}
+\left (1-\frac{1}{2\nabla}\right )^{2}\right ]-1, 
\label{eqn:RW} 
\ee 
where H$_{b}$ and H$_{0}$
are the pressure scale-heights at the bottom and top ends of the flux tube. We have compared
values of $\beta_{c}$ and $\beta_{c,RW}$ in Table 2, adopting appropriate values for the
quantities in Equation \ref{eqn:RW} from the stellar models used here (since $\nabla$, and
hence $\delta$, is assumed constant in the derivation of Equation \ref{eqn:RW}, a nominal value
of 2000 km is used as a representative extent of the flux tube).  Though the exact values of
$\beta_{c}$ and $\beta_{c,RW}$ do not match, owing to the reasons given above, they show
similar trends against T$_{eff}$.  It is evident that the weaker field strengths (i.e., larger
$\beta_{c,SI}$) for stability in cooler stars is mainly a reflection of smaller $\delta$.
Additionally, since the constant temperature gradient of a polytrope is proportional to $g$,
the scale-height ratio in Equation \ref{eqn:RW} yields small values ($\approx$ 1) at lower
values of log $g$ and hence large values of $\beta_{c}$.  We attribute the differences between
the present results and those from Equation \ref{eqn:RW} to the neglect of gradients in
$\nabla$, ionization and adiabatic index in the derivation of Equation \ref{eqn:RW}. It is
noted that since magnetic pressure varies as B$^{2}$ and the gas pressure is proportional to
$g$, difference between B$_{c}$ and B$_{eq}$ would be larger in higher gravity stars for a
constant value of $\beta_{c}$.

\section{Discussions and Conclusions}

We have examined the superadiabatic effect (or CC) \citep{parkr78} in the stellar context,
using realistic models of the outer convective layers of stars. Our results show that whereas
it is possible to produce highly evacuated stable tubes in stars with T$_{eff}\ge$ 5500 K
through the CC mechanism, it is not so in cooler stars; the decreasing amount of
superadiabaticity in the upper convection zone layers of K and M spectral type stars make the
CC inefficient yielding field strengths much less than B$_{eq}$. Hence, if CC is the main
physical process responsible for the formation of fibril-like evacuated flux tubes then it is
expected that the differences between the observed B and B$_{eq}$ increase as T$_{eff}$
decreases. The existing observational results are not entirely consistent with such a trend as
there are several cases of K and M dwarfs having B values close to or even exceeding B$_{eq}$;
but, all these stars with high B have high filling factors, $f$, typically 0.5 or higher, which
correlate strongly with stellar angular rotation frequency $\Omega$. Rotation-activity
correlation dominated through high $f$ values is a well observed fact \citep{saar90, saar96}
consistent with dynamo theory predictions. A careful look at the observational results compiled
by \citet{saar90} and \citet{solanki92} reveal that all the cases of B $<$ B$_{eq}$ also belong
to K and M spectral types, but which have smaller $\Omega$ and are less active; and all the G
type stars show B values tightly around B$_{eq}$. As an example, from the compilation by
\citet{saar90}, we find that the stars GL 171.2A (BY Dra) and HD 201091, which are of similar
spectral type (K5V, T$_{eff}$$\approx$4400 K), have widely differing field strengths: 2.8 kG
\citep{saaretal87} and 1.2 kG \citep{marcy-basri89} respectively. But these field strengths
have good correlation with the $\Omega$ and $f$ values (rotation periods of 1.85 and 37.9 days
and $f$ values of 0.5 and 0.24, respectively). From our results shown in Figure 1, we find for
T$_{eff}$$\approx$4400 K and a log $g$$\approx$4.5 (main-sequence) a field strength of
$\approx$ 1.3 kG that CC yields, in close agreement with the slow rotating, low $f$ value case
of HD 201091.  \citet{marcy-basri89} themselves caution that the separation of B and $f$ values
is uncertain and that the flux $f$B is larger than expected for HD 201091. But, their speculation 
that B remains at the inferred value while $f$ changes by 2 orders of magnitude, which is not
inconsistent with the observed time variation in the chromospheric emission, finds support from
the present result.  The best support for the present calculations comes from a multi-line
infra-red Zeeman analysis of $\epsilon$ Eridani, a K2 V star with T$_{eff}$=5130 K and log
$g$=4.7, by \citet{valentietal95}, who find $f$=0.088 and B=1.44 kG in close agreement with the
value B$_{c}\approx $ 1.4 kG from Figure 1.  It would thus seem that the CC indeed operates on
all solar-like main sequence stars producing convectively stable tubes as dictated by the
sub-surface superadiabatic structure; and, the higher than expected B values in K and M dwarfs
may originate from effects induced by high values of $\Omega$.  Recent detections of increased
photometric variability in highly active K dwarfs, which exhibit $saturation$ in their magnetic
activity \citep{O'delletal95}, indicate that stars with angular velocities $\Omega
>\Omega_{sat}$ show increased number of spots. Thus, the present result that B$_{c}$
$\ll$B$_{eq}$ for non-spot small scale magnetic fields in stars with T$_{eff}\le$5000 K and its
agreement with B observed on slow rotators provide a theoretical reason to believe the idea
that $\Omega_{sat}$ marks a change in contributions from $f$ and B to $f$B \citep{saar96b}: $f$
saturates around 0.6 and $\Omega$($ > \Omega_{sat}$) begins to contribute to B by increasing
$f_{spot}/f$ to maintain the continued increase of flux $f$B well described by the power-law
fits $f$B$\propto \Omega^{1.3}$ or $f$B$\propto (\tau_{c}\Omega)^{1.2}$ to the observations
\citep{saar90}, where $\tau_{c}$ is the convective turn-over time and $\tau_{c}\Omega$ is the
inverse Rossby number.  We suggest that the calculated stability limits B$_{c}$ can be used to
separate $f$ from the observed fluxes $f$B for slow rotators ($\Omega < \Omega_{sat}$), when
there remain otherwise unknown uncertainties in the separation of $f$ and B in observational
analyses. The calculations here extend down only to T$_{eff}$=4000 K due to limitations in
generating reliable convection zone models for cooler stars, but, by extrapolation, it is
likely that CC remains ineffective yielding B$_{c}(\tau_{c}=1)\ll $ B$_{eq}(\tau_{c}=1)$ in
cooler M dwarfs too.  Two similar spectral type M dwarfs, observed by \citet{krull-valenti96},
do show widely differing field strengths, in correlation with their $\Omega$, but the lower
$\Omega$ star seems to show B larger than the likely B$_{c}$. Future refinements, both in
observations and theory, are needed to understand fields in M dwarfs.

We note that the thin tube approximation does not get constrained by increasing $f$;  as long
as the distribution of flux remains in the form of small flux tubes dominating the observed
spectral profiles, the present results on B$_{c}$ can be used in the interpretation of
observations. Thus, if the observed profiles do arise only from non-spot fields even in fast
rotating K and M dwarfs which show B$\gg$B$_{c}$ then the present results call for new physical
mechanisms to produce highly evacuated tubes, which are likely not due to thermodynamic reasons
but due to fast-rotation induced phenomena. This may imply a failure of solar analogy as far as
the formation and dynamics of surface magnetic fields are concerned. We conclude by pointing
out that, in any case, the convective stability limits calculated here need to be satisfied by
pressure confined flux tubes and thus may serve as useful lower limits.

\acknowledgments
We are thankful to an anonymous referee for constructive comments.

\newpage

\begin{table}[ht]
\small
\footnotesize
\scriptsize
\caption{Characteristics of Model Atmospheres}
\label{tab:models}
\begin{tabular}{cccccccccccc}
\tableline
\multicolumn{12}{c}{log $g$} \\ \cline{2-12}
& \multicolumn{3}{c}{4.0} &  &\multicolumn{3}{c}{4.5} 
    				& & \multicolumn{3}{c}{5} \\
\cline{2-4}\cline{6-8}\cline{10-12} \\
T$_{eff}$ & Depth &$\delta_{max}$ &$\Gamma_{1}$&& Depth &$\delta_{max}$ &
$\Gamma_{1}$ && Depth &$\delta_{max}$ &$\Gamma_{1}$ \\ 
(K)&(km.)&&(at $\delta_{max}$)&&(km)&&(at $\delta_{max}$)&&(km)&&(at $\delta_{max}$)\\
\tableline
4000 & 3610 & 0.2306  &  1.665 &&1270 & 0.3173 &  1.663 &&295  & 0.0525 &  1.636 \\
&&(z=60 km)&&&&(z=30 km)&&&&(z=60 km)&\\
     & & & & & & & & \\
4500 & 4693 & 0.4606 &  1.590 && 1206 & 0.2596 &  1.593 &&423  & 0.1422 &  1.592 \\
&&(z=230 km)&&&&(z=80 km)&&&&(z=30 km)&\\
     & & & & & & & & \\
5000 & 5573 & 0.9000 &  1.553 && 1382 & 0.4735  &  1.544&&435  & 0.2450 &  1.515 \\
&&(z=180 km)&&&&(z=70 km)&&&&(z=30 km)&\\
     & & & & & & & & \\
5500 & 8300 & 1.2635 &  1.296 &&1604 & 0.7310 &  1.466 && 606  & 0.5004 &  1.446 \\
&&(z=130 km)&&&&(z=50 km)&&&&(z=20 km)&\\
     & & & & & & & & \\
6000 & 9030 & 1.627 &  1.369 &&2012 & 1.041 &  1.415 && 720  & 0.6028 &  1.501 \\
&&(z=80 km)&&&&(z=30 km)&&&&(z=10 km)&\\
     & & & & & & & & \\
6500 & 108487& 1.936 &  1.277 &&3472 & 1.318 &  1.341 && 960  & 0.8546 &  1.265 \\
&&(z=60 km)&&&&(z=20 km)&&&&(z=10 km)&\\
\end{tabular}
\end{table}

\newpage
\begin{table}[ht]
\caption{Comparison of Critical Values of $\beta$}
\label{tab:betas}
\begin{tabular}{ccccccccc}
\tableline
\multicolumn{9}{c}{log $g$} \\ \cline{2-9}
& \multicolumn{2}{c}{4.0} &  &\multicolumn{2}{c}{4.5} 
                                & & \multicolumn{2}{c}{5} \\
\cline{2-3}\cline{5-6}\cline{8-9} \\
T$_{eff}$ & $\beta_{c}$ &$\beta_{c,RW}$ &&$\beta_{c}$& $\beta_{c,RW}$ &&$\beta_{c}$ &
$\beta_{c,RW}$  \\ 
\tableline
4000 & 1.81 & 7.76 & &2.06&2.57&&14.44&6.90 \\
     & & & & & & & & \\
4500 & 1.19&7.32&&1.98&3.37&&3.94&2.16 \\
     & & & & & & & & \\
5000 &0.74&6.20&&1.13&2.54&&2.56&1.20 \\
     & & & & & & & & \\
5500 & 0.38&6.17&&0.52&2.35&&1.18&0.78 \\
     & & & & & & & & \\
6000 & 0.27 &6.44&&0.19&2.44&&0.54&0.86 \\
     & & & & & & & & \\
6500 & 0.12&6.95&&0.1&2.58&&0.16&0.76 \\
\end{tabular}
\end{table}

\newpage

\begin{figure}
\plotone{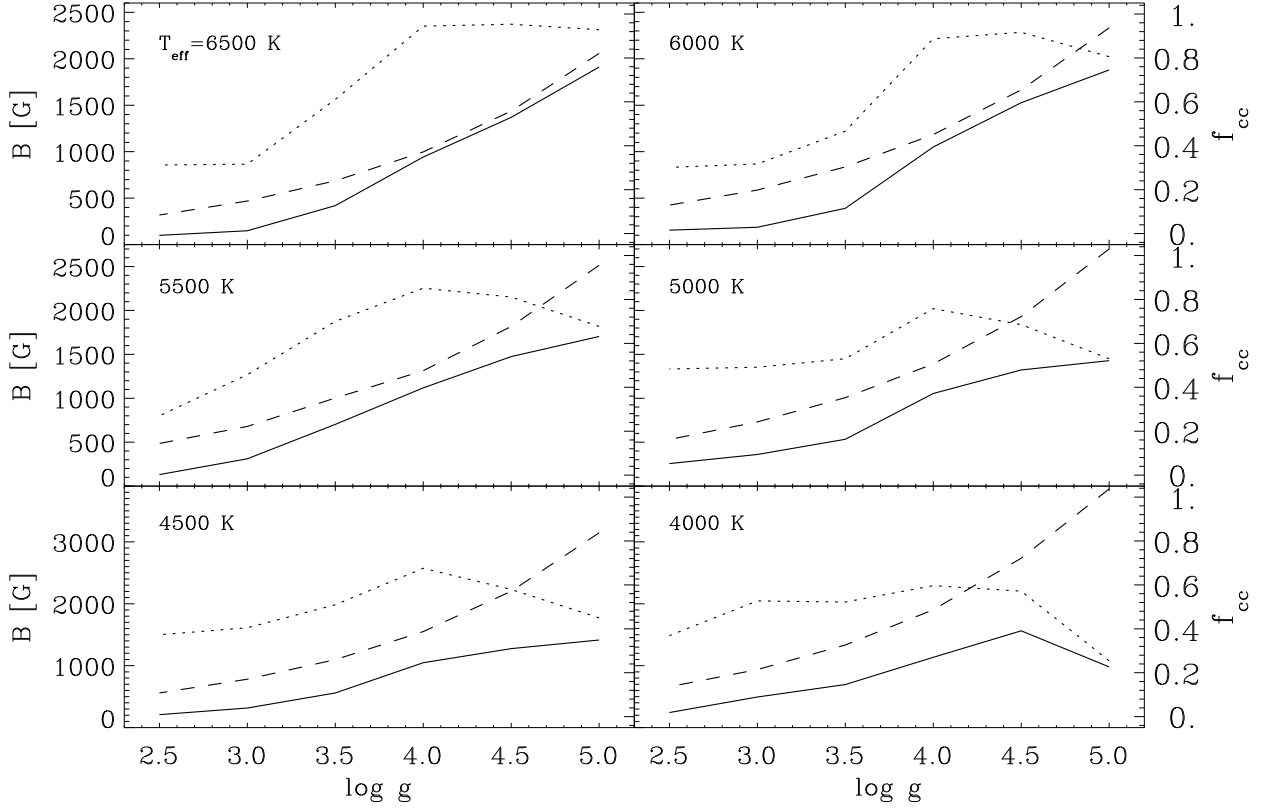}
\caption{The critical magnetic field strengths B$_{c}(\tau_{c}=1)$ for convective stability
($solid$ $lines$) and thermal equipartition limits B$_{eq}(\tau_{c}=1)$ ($dashed$ $lines$),
against log $g$ for various T$_{eff}$. B$_{c}$
correspond to those that can be achieved by convective collapse and B$_{eq}$
are the maximum possible field strengths for flux tubes. Efficiencies of CC, $f_{cc}$,
are shown as $dotted$ curves and correspond to right-side ordinates of the panels.}
\label{fig:loggB}
\end{figure}

\begin{figure}
\plotone{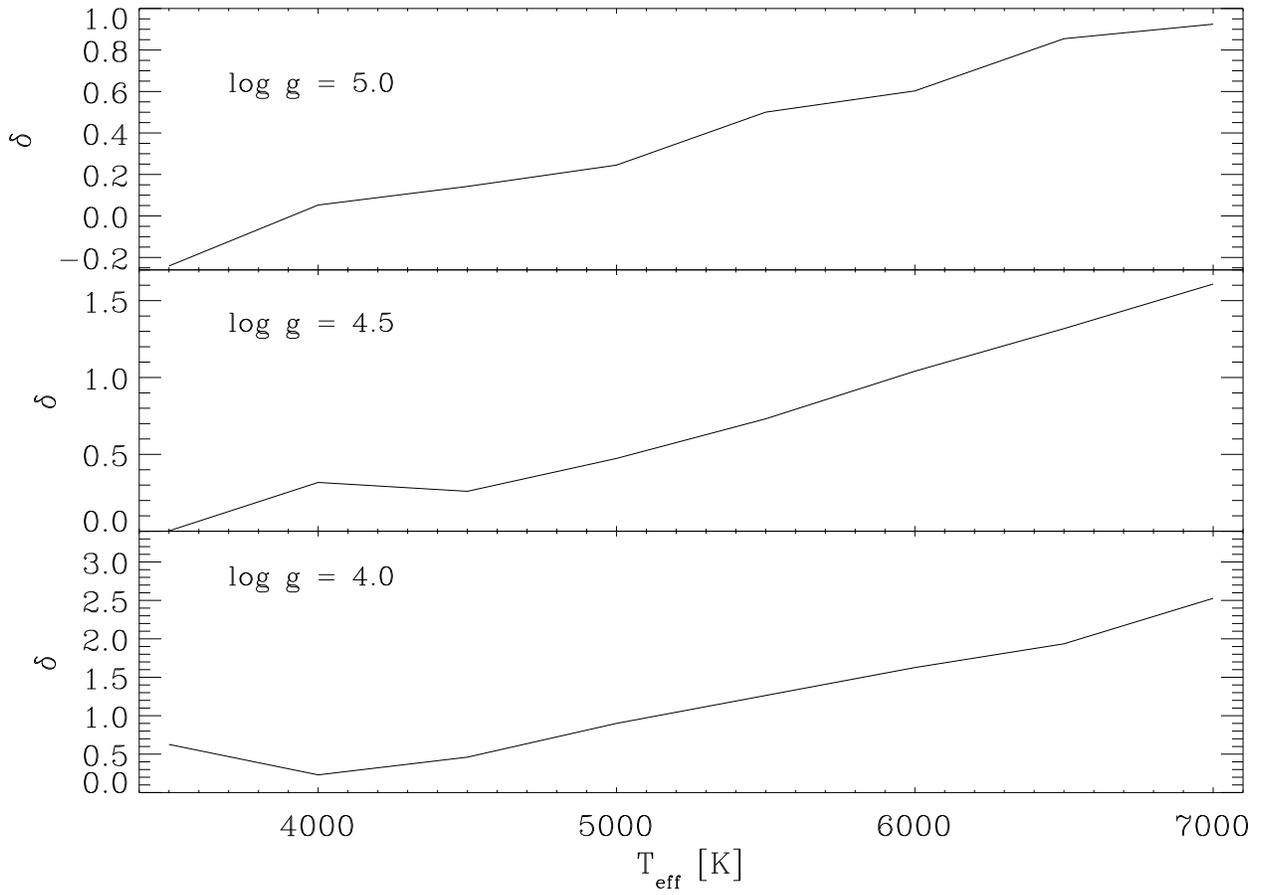}
\caption{The peak values of superadiabaticity $\delta$=$\nabla$-$\nabla_{a}$ that
characterize the sub-surface upper convective layers as a function of T$_{eff}$ for
three different values of log $g$.}
\label{fig:supad}
\end{figure}
\end{document}